\documentstyle{article}
\newcommand{\beq}{\begin{equation}}
\newcommand{\eeq}{\end{equation}}
\newcommand{\mc}{m_{\chi}}
\newcommand{\sv}{<\sigma V>}
\newcommand{\bino}{\tilde{{\rm B}}}
\newcommand{\higgsino}{\tilde{{\rm H}}}
\newcommand{\msf}{m_{\tilde{{\rm f}}}}

\title{{\bf {\sc Searching for TeV dark matter by atmospheric \u{C}erenkov
techniques}}}

\author{M. Urban\thanks{Laboratoire de Physique Nucl\'eaire et Hautes Energies,
Ecole Polytechnique/CNRS-IN2P3, 91128  Palaiseau Cedex , FRANCE} \and A.
Bouquet\thanks{LPTHE Universit\'es PARIS 6 and 7, Unit\'e associ\'ee au CNRS UA
280, 2 Place Jussieu, 75251 Paris Cedex 05 , FRANCE} \and B.Degrange$^*$ \and
P.
Fleury$^*$ \and J. Kaplan$^{\dagger}$ \and A.L. Melchior$^{\dagger}$ and E.
Par\'e$^*$}
\date{}

\begin{document}
\pagestyle{empty}
\vspace{4cm}
\maketitle
\sloppy
\vspace{2cm}

\begin{abstract}
\vspace{1cm}
{\large There is a growing interest in the possibility that dark matter could
be
formed of  weakly interacting particles with a mass in the 100 GeV - 2 TeV
range, and supersymmetric  particles are favorite candidates. If they
constitute the dark halo of our Galaxy, their mutual  annihilations produce
energetic gamma rays that could be detected using existing atmospheric
\u{C}erenkov techniques.}
\end{abstract}

\vspace{4cm}
\noindent {\em To appear in Physics Letters B}
\vspace{2 cm}
July 1992
\vspace{2cm}

\newpage
\pagestyle{plain}

\section{Introduction}

Galaxies seem embedded in massive spheroidal dark halos
\cite{faber79,trimble87,kormandy87,turner90}. Extensions of the standard model
of particle  physics often predict the existence of a new stable particle, a
natural candidate for halo dark matter. The residual  abundance of a particle
after the Big-Bang depends on its mass and interactions, and to account for the
dark matter  puzzle a weakly interacting massive particle (W.I.M.P.) is needed,
for instance a neutrino-like particle with a  mass around 1 TeV. Such particles
will be very difficult to detect, even by accelerator experiments (LHC-SSC)  of
next century. Halo particles can be directly detected when they cross a
bolometer detector \cite{primack88,smith89} but the flux is  low above 50 GeV.
They can also be indirectly detected through their annihilation products
\cite{gunn78,stecker78,silk84,bergstrom88,rudaz89,bouquet89}~: annihilations
after their capture in the Earth or in the Sun producing neutrinos, or
annihilations in the halo producing energetic  cosmic rays (gamma rays,
electrons, protons, antiprotons$\ldots$). Existing models for our galactic halo
predict a  strong annihilation rate near the center of the Galaxy. We focus
here on the annihilations into photons, because  photons conserve their initial
direction. Each annihilation produces either primary photons (leading to a
gamma  line signal) or secondary photons from the decay of primary annihilation
products. The fluxes are low again for a  large WIMP mass, but this can be
balanced by the large acceptance of detectors using the atmospheric
\u{C}erenkov  technique like the existing Atmospheric \u{C}erenkov Telescope
(ACT) of the Whipple Observatory \cite{weekes88}. With similar instruments, it
should be possible to detect for instance the leading WIMP candidate, the
neutralino predicted by supersymmetry \cite{nilles84}.

\section{ACT's and backgrounds}

The ACT detects incoming cosmic rays through the \u{C}erenkov light emitted by
the shower of secondary particles created when the cosmic ray (photon,
electron, proton, heavier nucleus$\ldots$) collides with a nucleus in the upper
atmosphere \cite{weekes88}. The image shape analysis affords an efficient gamma
versus
hadron separation. The \u{C}erenkov light covers a disk of about 120 m
radius on the Earth surface (an area of 45000 m$^2$), and the detector
must be somewhere in this area to see the cosmic ray shower. Less energetic
cosmic rays produce less \u{C}erenkov light, and the minimal detectable energy
(threshold) is linked to the size of the mirrors which collect this light. The
best observation conditions correspond to a cosmic ray within 45 degrees from
the zenith. As we shall see, the photon flux from WIMP annihilations is largest
from the galactic center, which can be seen at zenith only if the observatory
is built in the southern hemisphere, at a latitude near -29 degrees.
Observation of \u{C}erenkov light is only possible at night when there is no
moon, and so the galactic center can be observed for $t_{obs} \simeq 2 \times
10^6$ s per year. The effective angular acceptance $\Delta \Omega$ of
present ACT's is about $10^{-3}$ sr with an angular resolution of $3 \times
10^{-6}$ sr, so that they are well suited to the study of localized
sources. The usual procedure consists in the comparison of the flux from an
expected source with neighbouring sky regions (on-off experiment).

To be definite, we shall use as a reference ACT the Granite set-up at the
Whipple Observatory, which uses two mirrors of about 10 m diameter, 120 m
apart, leading at present to a minimal detectable energy around 200 GeV.
Increasing the number and/or the size of mirrors may lower this threshold. The
two telescopes effectively survey 63000 ${\rm m^2}$ if they are used
independantly, but their effective detection area $S$ is only 18000 ${\rm
m^{2}}$ when the two telescopes are used in coincidence to give the best
energy resolution and an increased hadronic rejection. The energy of a gamma
cosmic ray is determined from a stereoscopic view with a precision $\sigma
(E)/E \simeq 0.1$ . Adding other telescopes, about 120 m apart, would
increase the effective area and therefore lower the detectable cosmic flux.
The product $S \Delta \Omega t_{obs}$ is a measure of the efficiency of the
telescope to detect cosmic rays, and $S \Delta \Omega t_{obs} = 3.6 \times
10^{11}$ cm$^2$ sr s for our reference ACT.

There is a large background to the WIMP annihilation signal that we are
looking for. Cosmic ray hadrons must be rejected, and experience at the
Whipple Observatory shows that a rejection factor of 10$^3$ should be reached
for two detectors in coincidence. The flux of misidentified hadrons then
follows a power law \cite{weekes88}~:
\beq
\frac{{\rm d}N_h}{{\rm d}t \, {\rm d}S \, {\rm d}\Omega \, {\rm d}E} = 6 \times
10^{-9} \left( \frac{E}{1 \, {\rm TeV}} \right)^{-2.75} \; {\rm photon-like \,
hadrons \, (s \, cm^2 \, sr \, TeV)^{-1} }
\label{fluxhadr}
\eeq
When a hadron is misidentified as a photon, its energy is underestimated, and
the preceding formula takes this effect into account. Electrons cannot be
distinguished from photons with atmospheric \u{C}erenkov detectors, as their
showers have the same development as photons in the atmosphere. The electron
flux is measured at energies up to 1 TeV \cite{nishimura80,he91}. The spectrum
also follows a power law~:
\beq
\frac{{\rm d}N_e}{{\rm d}t \, {\rm d}S \, {\rm d}\Omega \, {\rm d}E} = 6 \times
10^{-9} \left( \frac{E}{1 \, {\rm TeV}} \right)^{-3.3} \; {\rm electrons \, (s
\, cm^2 \, sr \, TeV)^{-1} }
\label{fluxelec}
\eeq
The electron background dominates the background of misidentified hadrons up to
1
TeV. The fluxes of electrons or hadrons are isotropic. Finally, there is the
diffuse photon background itself, supposed to be due to the collisions of
cosmic
ray protons on interstellar gas. The spectrum from the galactic center was
measured by the COS-B satellite \cite{mayer82,bhat85,bloemen89} up to 5 GeV~:
\beq
\frac{{\rm d}N_{\gamma}}{{\rm d}t \, {\rm d}S \, {\rm d}\Omega \, {\rm d}E} =
0.35 \times 10^{-9}
\left( \frac{E}{1 \, {\rm TeV}} \right)^{-2.75} \; {\rm photons \, (s \, cm^2
\,
sr \, TeV)^{-1} }
\label{fluxgam}
\eeq
This power law is the same as for the hadron spectrum, and we may assume that
Equation~\ref{fluxgam} extends to higher energies. This is not in disagreement
with upper bounds on the photon flux above 5 GeV \cite{nishimura80,he91}. This
flux is  smaller than the electron background.

A signal can be detected even if it is small compared to the background,
provided that the background is measured with a good precision. In our case,
an on source-off source method will be available, allowing for a reliable
estimation of this background. The noise is the uncertainty on the background,
that is the square root of the number $N_{backgr}$ of all photon-like events
(including electrons and misidentified hadrons) detected during the experiment
time $t_{obs}$ in a given energy bin $\Delta E$, that we shall take as $3
\sigma (E) \simeq 0.3 E$. For a detector of acceptance $S \Delta \Omega$, this
number is~:
\beq
N_{backgr} = S \, \Delta \Omega \, t_{obs} \left \{ \, \frac{{\rm
d}N_{\gamma}}{{\rm d}t \, {\rm d}S \, {\rm d}\Omega \, {\rm d}E} \, + \,
\frac{{\rm d}N_e}{{\rm d}t \, {\rm d}S \, {\rm d}\Omega \, {\rm d}E} \, +
\,\frac{{\rm d}N_h}{{\rm d}t \, {\rm d}S \, {\rm d}\Omega \, {\rm d}E} \,
\right \} \, \Delta E
\label{backgr}
\eeq
We must now compare these numbers to the expected signal flux from halo WIMP
annihilations.

\section{WIMP annihilation from the galactic center}

We assume that our Galaxy is embedded in a dark halo made of WIMP's of mass
$\mc$ which annihilate in pairs. The annihilation rate is
$\sv_{halo} n_{\chi}^{2}$, where $\sv_{halo}$ is the (thermal average in the
halo of the) annihilation cross-section $\sigma$ times the relative WIMP
velocity $V$, $n_{\chi} = \rho_{\chi}/\mc$ is the dark matter number density.
At large distance $r$ from the center, the halo density decreases as~:
\beq
\rho (r) \simeq \rho_{\odot} \, \frac{a^2 + r_{\odot}^2}{a^2 + r^2}
\eeq
where $\rho_{\odot} = 0.3 \pm 0.1 \, {\rm GeV/cm^3}$ is the dark matter density
in the solar neighbourhood, $a = 5 \pm 3$ kpc is the core radius and $r_{\odot}
= 8.5$ kpc is the distance of the Sun to the galactic center
\cite{bahcall80,caldwell81,binney87,flores88}. The annihilation rate of this
dark matter far away from the galactic center turns out to be too low to be
detected with present size ACT's.

The dark matter density may however be much larger at the center, as there
seems
to be a massive nucleus at the center of the Galaxy
\cite{zeldovich80,clemens85}. According to the isothermal model of Ipser
and Sikivie \cite{ipser87}, this nucleus induces an enhancement of the dark
matter density within a 150 pc radius around the center, with a density at
center of~:
\beq
\rho \simeq 0.34 \; \alpha \; {\rm TeV/cm^3}
\eeq
The $\alpha$ parameter ($0.1<\alpha<3$) reflects the uncertainties on the
rotation velocity of the galaxy, and on the contribution of luminous matter to
this velocity. Annihilation photons in the TeV energy range are unlikely to be
absorbed by  baryonic matter in the galactic center since, in the worst case of
a completely gaseous core, a $\gamma$-ray crosses less than one radiation
length. As seen from Earth, the central region covers $\Delta \Omega =
10^{-3}$ sr, nicely corresponding to the acceptance of present ACT's. We can
perform an on-off experiment, aiming at the galactic center and comparing with
the result off center. The $10^{-6}$ sr ACT angular resolution is small enough
to distinguish point sources inside these $10^{-3}$ sr from the more diffuse
WIMP annihilation cloud.

Photons appear in the decay chain of the primary annihilation products (quarks,
leptons, Higgs scalars, W$^\pm$ or Z$^0$'s, etc.), and their energy is then
continuously distributed. Their distribution ${\rm d}n_{\gamma}(E)/{\rm d}E$
can be
computed through a Lund Monte-Carlo \cite{bengtsson87} (Figure 1), and depends
only weakly on the WIMP model. The integral $\int {\rm d}n_{\gamma}(E)$ is the
photon multiplicity per annihilation. The number of photons due to WIMP
annihilations within 150 pc  from the center, and received on our reference
telescope is~:
\beq
N_{\gamma}(E) \,{\rm d}E = S \,t_{obs} {\rm d}n_{\gamma}(E) \,
\frac{\sv_{halo}}{4 \pi r^2_{\odot}} \, \int_0^{150 \, {\rm pc}} \left(
\frac{\rho(r)}{\mc} \right) ^2 4 \pi r^2 {\rm d}r
\label{flux}
\eeq
WIMP annihilations outside the 150 pc sphere, but inside the $10^{-3}$ sr
cone, do not contribute to a signal defined by an on-source minus off-source
subtraction. The cross-sections $\sigma$ and the corresponding energy spectra
${\rm d}n_{\gamma}(E)/{\rm d}E$ , reduced to a delta function in the case of
monochromatic
photons, are summed over all relevant annihilation channels.

Two different signals can be looked for at the same time. We can look for a
gamma line due to the direct annihilation of WIMP's into photons, or we can
look for the total excess of soft photons due to the decays of all
annihilation products.

{\bf Soft photons}

The Lund Monte-Carlo codes \cite{bengtsson87} typically indicate that 50
photons or so are produced in the decay chains, but  that most of them are at a
very low energy compared to the WIMP mass $\mc$ (see Figure 1). The signal is
the number of photons received above the ACT energy threshold $E_{thresh}$.
Table 1 shows the integral $\int {\rm d}n_{\gamma}(E)$ from $E_{thresh}$ to
$\mc$ as a function of $\mc$ for a quark-antiquark channel (the result being
very similar for other channels).

\begin{center}
\begin{tabular}{|l|r|r|r|r|r|}
\hline
$\mc$ (TeV) & 0.4 & 0.8 & 1.0 & 2.0 & 4.0 \\ \hline
$E_{thresh} = 0.1$ TeV & 0.02 & 0.17 & 0.32 & 1.45 & 4.40 \\ \hline
$E_{thresh} = 0.2$, TeV & 0.0004 & 0.02 & 0.05 & 0.35 & 1.55 \\ \hline
\end{tabular}

{\bf Table 1}~: Number of soft decay photons above threshold. The first line
gives the mass $\mc$ of the particles which annihilate into a quark-antiquark
pair, the second line gives the number of decay photons with an energy above
0.1 TeV, and the third line the number of photons above 0.2 TeV.
\end{center}

Note the great sensitivity to the threshold~: the number of photons above the
treshold increases roughly as $\mc^2$, and, as a result, the signal can only
be detected if the WIMP mass is far above threshold. From Equation~\ref{flux},
and for our reference telescope, this signal is~:
\beq
{\rm Soft \, photon \, signal} = \int_{E_{thresh}}^{\mc} N_{\gamma}(E) \, {\rm
d}E = 8300 \,{\rm photons/year} \, \alpha^2 \int {\rm d}n_{\gamma}(E) \,
\frac{\sv_{halo}}{10^{-26}\, {\rm cm^3/s}} \left(
\frac{1 \, {\rm TeV}}{\mc}\right) ^2
\label{soft}
\eeq
The corresponding noise is the square root of the total number of background
photons received above $E_{thresh}$. From
Equations~\ref{fluxhadr}, \ref{fluxelec} and \ref{fluxgam}, it is
approximately~:  \beq
{\rm Soft \, photon \, noise} = \left\{\int_{E_{thresh}}^{\mc} N_{backgr}(E) \,
{\rm d}E \right\}^{1/2} \simeq 220 \,{\rm photons/\sqrt{year}} \;
\frac{0.2 \, {\rm TeV}}{E_{thresh}}
\eeq
Note that we are conservative in integrating up to $\mc$ because the integral
$\int {\rm d}n_{\gamma}(E)$ is saturated well before $\mc$, and therefore we
are adding noise for no
signal.

The signal to noise ratio will be very low below $\mc \simeq 1$ TeV
unless annihilation cross-sections are large. But they cannot be very far from
$\sv_{halo} \simeq 10^{-26}\, {\rm cm^3/s}$ , if WIMP's constitute the dark
matter of the universe. WIMP's were produced in large numbers during the hot
dense phase of the Big-Bang, and to constitute the dark matter, some fraction
must have survived annihilation up to now. This annihilation becomes negligible
when the temperature gets lower than a critical decoupling temperature, and
the present relic density $\Omega_{\chi}$ (in unit of the critical density) is
\cite{kolb90}~:
\beq
\Omega_{\chi}h^2 = \frac{2.5 \times 10^{-27} \, {\rm cm^3/s}}{\sv_{BB}}
\label{omega}
\eeq
up to logarithmic corrections ($h$ is the Hubble constant divided by 100
km/s/Mpc, and we shall take $h = 0.5$). Note that the thermal average
$\sv_{BB}$ of the annihilation cross-section at the Big-Bang can differ from
the thermal average $\sv_{halo}$ in the halo, since the temperature at
decoupling differs from the halo equivalent temperature.

To account for dark matter in galactic halos, a mean density $\Omega_{halo}
\simeq 0.1$ is required \cite{faber79,trimble87,kormandy87,turner90}, while a
mean density $\Omega \simeq 1$ seems required on larger scales. When the WIMP
annihilation rate is strong, $\Omega_{\chi}<<1$, and the density of WIMP's in
the halo cannot be much larger than $\rho_{\chi} \simeq \rho_{halo}
\Omega_{\chi}/\Omega_{halo}$ . The remaining component could be due to brown
dwarfs \cite{rees86} or to another WIMP. To interpolate the halo WIMP density
between this strong annihilation rate and the weak annihilation rate (where
$\rho_{\chi} \simeq \rho_{halo}$), Griest, Kamionkowski and Turner (GKT)
\cite{griest90} suggested the formula $\rho_{\chi} \simeq \rho_{halo} /[1 +
\Omega_{halo}/\Omega_{\chi}]$ , but any smooth interpolation could be chosen.
This GKT correction decreases the expected flux for a large annihilation
cross-section, and Equation~\ref{soft} becomes~:
\beq
{\rm Soft \, photon \, signal} \simeq 8300 \,{\rm photons/year} \, \alpha^2
\int
{\rm d}n_{\gamma}(E) \, \frac{\sv_{halo}}{10^{-26}\, {\rm cm^3/s}} \left(
\frac{1 \, {\rm TeV}}{\mc}\right) ^2 \frac{1}{[1 \,+\,
\Omega_{halo}/\Omega_{\chi}]^2} \eeq

{\bf ''Line'' signal}

Two WIMP's can directly annihilate into two photons ($\int {\rm d}n_{\gamma}(E)
= 2$), and the energy of the photons is then equal to the mass $\mc$ of the
(non-relativistic) WIMP, leading to a gamma line. A narrow line in the range
$10^{2\pm1}$ GeV would provide a clear signature for dark matter annihilation
\cite{gunn78,stecker78,silk84,bergstrom88,rudaz89,bouquet89}, but the
cross-section $\sigma_{2\gamma}$ for annihilation into 2 photons strongly
depends on the WIMP model, and moreover it is expected to be much smaller than
the annihilation cross-sections into a pair of quarks, leptons, Higgs or gauge
bosons. For our reference telescope, the expected number of annihilation
photons in the line at $E = \mc$ is~:
\beq
{\rm Line \, signal} = 166 \,{\rm photons/year} \, \alpha^2 \,
\frac{<\sigma_{2\gamma}V>_{halo}}{10^{-28}\, {\rm cm^3/s}} \left( \frac{1 \,
{\rm TeV}}{\mc}\right) ^2 \frac{1}{[1 \,+\, \Omega_{halo}/\Omega_{\chi}]^2}
\eeq
 From Equation~\ref{backgr}, the background noise for such a $\gamma$ line,
with
an energy bin $\Delta E = 3 \,\sigma = 0.3\, E$, is well approximated by~:
\beq
{\rm Line \; noise} = \sqrt{N_{backgr}} = 35 \,{\rm photons/\sqrt{year}} \;
\frac{1 \, {\rm TeV}}{\mc}
\eeq
Annihilations into a photon and another particle of mass $M$ also gives a
monochromatic photon, at an energy $E \simeq \mc - M^2/4\mc$ . Some kind of
spectroscopy is thus in principle possible, depending on the energy resolution.

\section{Neutralinos}

We now focus on supersymmetry \cite{nilles84}, the favorite extension of the
standard model, because the lightest supersymmetric particle (LSP) $\chi$ is
stable due to a conserved quantum number, R-parity, and because its relic
density is naturally $\Omega_{\chi} \simeq 1$, making it a good dark matter
candidate. There are two neutral gauge fermions $\bino$ and $\tilde{{\rm W}}^3$
and two neutral Higgs fermions $\higgsino^1$ and $\higgsino^2$
which mix when gauge symmetry is broken, and the four resulting mass
eigenstates are called neutralinos. The lightest one, {\em the} neutralino,
usually is the LSP. Its mass  and couplings depend on 4 parameters~: the masses
$M_1$ and $M_2$ of the gauge fermions $\bino$ and  $\tilde{{\rm W}}^3$ before
gauge symmetry breaking (usually related by the GUT relation $M_1 = \frac{5}{3}
M_2 \tan^2\theta_W$), the Higgs fermion mass $\mu$ and the ratio $\tan\beta$ of
the vacuum expectation values of the two Higgs scalars
\cite{nilles84,griest90,ellis84,haber85,griest88,olive89}. In addition, the
neutralino cross-sections depend on the masses of the quarks and leptons, and
of their scalar partners, and on the Higgs bosons masses. There are weak direct
experimental lower bounds from LEP on the neutralino mass \cite{decamps90} but,
assuming a GUT relation between the gluino mass $M_3$ and the masses $M_1$ and
$M_2$, more stringent indirect bounds can be obtained from UA2 \cite{alitti90}
and CDF \cite{abe89} experiments. It is then likely that the mass of the
neutralino is larger than about 50 GeV.

We are interested in the possibility that the neutralino be heavier than 100 to
200 GeV, the ACT threshold, because it allows a neutralino relic density near
the critical density. We focus in this paper on the two extreme cases of a pure
bino or a pure higgsino. A mixed state is not a priori excluded, but is quite
unlikely in this high mass region~: when $|\mu| \gg M_2$, the lightest
neutralino is an almost pure $\bino$ state, and an almost pure $\higgsino$ when
$|\mu| \ll M_2$. The equations for annihilation cross-sections are extremely
complex \cite{griest90,ellis84,haber85,griest88,olive89}, and we only give here
a  sketch of the results.

Most of the time, a bino pair annihilates into a quark or a lepton pair,
through the exchange of the corresponding s-quark or s-lepton of mass $\msf$
($\msf > \mc$ since, by assumption, $\chi$ is the LSP). The thermally averaged
cross-section for binos at temperature $T$ is approximately~:
\beq
\sv = \frac{\pi \alpha^2}{216 \cos^4\theta_W} \, \frac{\mc^2}{(\msf^2+\mc^2)^2}
\left\{ 289 \frac{m_{top}^2}{\mc^2} \,+\, 9160 \frac{T}{\mc}
\frac{\msf^4+\mc^4}{(\msf^2+\mc^2)^2} \right\}
\label{sigbino}
\eeq
In this equation, $\alpha$ is the fine structure constant. As illustrated by
Equation~\ref{sigbino}, thermally averaged cross-sections for neutralino
annihilation can be written as  $\sv = a + bT/\mc$ . Temperatures are very low
in the galactic halo and the $a$ term dominates, whereas the $b$ term dominates
during the Big-Bang, down to the decoupling temperature $T_{dec} \simeq
\mc/25$.
Equations~\ref{omega} and \ref{sigbino} then yield a relation between the bino
mass and its relic density $\Omega_{\chi}$ ~:
\beq
\mc = 0.56 \, {\rm TeV} \, h \sqrt{\Omega_{\chi}}
\eeq
This value was obtained assuming that $\msf = \mc$, and it becomes much smaller
if $\msf \gg \mc$ . Therefore cosmology leads to an {\em upper} bound on the
bino mass \cite{griest90,ellis84,haber85,griest88,olive89}. It is very
interesting that a similar upper bound is also indicated by naturalness
arguments in the context of low energy supergravity models \cite{barbieri88}. A
negligibly small relic density is excluded for a bino by the experimental lower
bounds on the neutralino mass and binos should be around us in large numbers,
even if they do not solve the dark matter problem. Assuming again the lightest
possible s-top mass, namely $\msf = \mc$ , the cross-section for annihilation
in the halo into a top-antitop pair (for $\mc > m_{top}$) is~:
\beq
\sv_{halo} \, \simeq \, 2.8\times10^{-29} \, {\rm cm^3/s} \left(
\frac{m_{top}}{150 \, {\rm GeV}}\right) ^2 \, \left( \frac{1 \, {\rm TeV}}{\mc}
\right) ^4
\eeq
if the bino is heavier than the top quark. The top quarks decay into a large
number of soft photons, and protons, antiprotons, electrons, positrons,
neutrinos and antineutrinos. The photon energy spectrum ${\rm d}n_{\gamma}/{\rm
d}E$ is computed by Lund Monte-Carlo \cite{bengtsson87}. Figure 2 shows the
expected excess in the number of photons in the direction of the center of the
galaxy, seen by our reference telescope, as a function of the neutralino mass
$\mc$ . The excess number of photons is compared to the background noise
measured in directions away from the center of the Galaxy. The bino
annihilation into soft photons will be difficult to detect, unless the dark
matter density  enhancement at the galactic center is larger than we assumed
(i.e. if $\alpha \gg 1$).

Let us now turn to a $\higgsino$-like neutralino. The dominant higgsino
annihilation channels are into WW or ZZ pairs, through the exchange of a
chargino or a neutralino. The expression for its annihilation cross-sections is
very complicated \cite{griest90,ellis84,haber85,griest88}, but, in the
mass range we are interested in, it scales roughly as~:
\beq
\sv_{BB} \, \simeq \, 2.2\times10^{-26} \, {\rm cm^3/s} \, \left( \frac{1 \,
{\rm TeV}}{\mc} \right) ^2
\eeq
leading to~:
\beq
\mc \, \simeq \, 3 \, {\rm TeV} \, h \sqrt{\Omega_{\chi}}
\eeq
The bound is larger than for the bino, because higgsinos are more strongly
coupled than binos. Moreover, at low temperature, the higgsino cross-section
scales as $1/\mc^2$ (fermion exchange) while the bino cross-section scales
as $m_{top}^2/\mc^4$ (scalar exchange). In the halo, the cross-section for
higgsino pair annihilation into W's and Z's is~:
\beq
\sv_{halo} \, \simeq \, \frac{\pi \alpha^2}{16} \; \frac{2\cos^4\theta_W +
1}{\sin^4\theta_W \cos^4\theta_W} \; \frac{1}{\mc^2}
\eeq
Higgsino annihilation is larger than bino annihilation, both in the Big-Bang
(hence a smaller relic density for a given mass) and in the halo (hence a
larger photon flux). Figure 2 indeed shows that higgsino annihilation into
soft photons would be easier to detect than bino annihilation. However the
signal is still marginal for our reference telescope.

A gamma line would be a clearer signal . A neutralino pair cannot annihilate
into two photons at the tree level, one loop graphs are required and probably
lead to a tiny cross-section. There are many relevant graphs, in particular if
annihilations into one photon and another particle (Z$^0$, Higgs scalar) are
included. We are computing these graphs, but this long and tedious work is not
yet completed. Nevertheless, an order of magnitude estimate may be derived
from the graphs already computed \cite{giudice89}. For the annihilation of a
bino into two photons through quark/s-quark or lepton/s-lepton loops, this
leads to~:
\beq
<\sigma_{2\gamma}V>_{halo} \; \simeq \; 4.2\times10^{-29} \, {\rm cm^3/s} \,
\left( \frac{1 \, {\rm TeV}}{\mc}\right) ^2 \, \left( \frac{\mc}{\msf} \right)
^4
\label{sig2gammas}
\eeq
Of course, cancellations may happen between graphs, and the cross-section of
Equation~\ref{sig2gammas} may be grossly  overestimated. If it is not the case,
Figure 3 shows that a bino could be detected at the 5 standard deviation level
within one year of observation, down to the present 0.2 TeV threshold, again
assuming $\msf = \mc$. The cross-section for the annihilation of a higgsino
into two photons is not yet known, and we boldly assumed the same
cross-section as for a bino. Figure 3 shows that the GKT correction factor
suppresses the higgsino line signal, due to the smaller higgsino relic density
for a given mass. Thus a higgsino would be more difficult to detect in this
way. The two searches, for a soft photon excess and for a line signal, are
therefore complementary, as they are sensitive to different types of
neutralinos, and they can be done simultaneously within the same experiment.

\section{Conclusions}

Dark matter made of particles in the 100 GeV - 1 TeV mass region could be
detected through its annihilation products by an atmospheric \u{C}erenkov
telescope searching for energetic gamma rays. We apply this idea to the
neutralino WIMP candidate from supersymmetry, and show that it may work,
provided that~:

 i) the neutralino is heavier than about 200 GeV, which is one of the windows
allowing a relic density $\Omega_{\chi}$ larger than 10$^{-2}$.

 ii) the density enhancement in the center of the Galaxy is as large as
expected by the isothermal Ipser-Sikivie model.

iii) a large enough atmospheric \u{C}erenkov detector is built in the southern
hemisphere, and aimed towards the galactic center.

Improvements of the ACT technique to lower the energy threshold are of
paramount importance to close the gap between the 0.05 TeV or so of present
accelerator experiments, and the 0.2 TeV expected for our reference
telescope. A complementary study could be based on the detection of the charged
decay products, the electrons, protons and antiprotons. There is no
directionality any more, and so no on-off experiment is possible, but
consequently there is no prefered location for the apparatus. Integrating all
over the galactic halo, preliminary calculations suggest that antiprotons (and
protons) must display a strong enhancement below the neutralino mass, while
other nuclei should not.

{\Large {\bf FIGURE CAPTIONS}}

Figure 1 ~: Photon energy spectrum from the pair annihilation of neutralinos,
showing the continuum due to quarks, leptons, Higgs and gauge boson decays,
computed through a Lund Monte-Carlo, and the lines (convolved with
experimental energy resolution) due to direct annihilations into two photons,
and a heavy particle and a photon. The relative size of the lines and of the
continuum is arbitrary.

Figure 2 ~: Soft photon signal. Expected excess number
of events per year above a 100 GeV threshold for a bino or a higgsino-like
neutralino, as a function of the neutralino mass $\mc$, compared to the noise
(one-sigma fluctuations of the background). The bino and higgsino curves
correspond to an Ipser-Sikivie model of the halo core with $\alpha = 1$, using
the GKT correction to reduce the flux when the neutralino relic density
$\Omega_{\chi}$ is smaller than the halo mean density $\Omega_{halo} = 0.1$ ,
and assuming a s-fermion mass $\msf = \mc$. The two vertical lines correspond
to $\Omega_{\chi} h^2 = 1$ for a bino and for a higgsino.

Figure 3 ~: Line signal. Expected number of events per year in the line for a
bino or higgsino-like neutralino, as a function of the neutralino mass $\mc$,
compared to the noise (one-sigma fluctuations of the background). The bino and
higgsino curves correspond to an Ipser-Sikivie model of the halo core with
$\alpha = 1$, using the GKT correction, and assuming a s-fermion mass $\msf =
\mc$, and an energy resolution $\Delta E = 0.3 \,E$. The two vertical lines
correspond to $\Omega_{\chi} h^2 = 1$ for a bino and for a higgsino.


\begin{thebibliography}{99}

\bibitem{faber79}	S.M. Faber and J.S. Gallagher, Ann. Rev. Astron. Astrophys.
17
(1979) 135

\bibitem{trimble87}	V. Trimble, Ann. Rev. Astron. Astrophys. 25(1987)425

\bibitem{kormandy87}	J. Kormandy and G.R. Knapp (eds), Proceedings of the IAU
Symposium 117, ''Dark matter in the universe'' (Reidel 1987)

\bibitem{turner90}	M.S. Turner,''Dark matter in the universe'', talk at Nobel
Symposium 79 ''The birth and early evolution of our universe'' (Graftavallen,
11-16 June 1990) Fermilab-Conf 90/230A

\bibitem{primack88}	J. Primack, D. Seckel and B. Sadoulet, Ann Rev. Nucl. Part.
Phys. 38(1988)751

\bibitem{smith89}	P.F. Smith and J.D. Lewin, in 3rd ESO-CERN
Symposium, M. Caffo et al. eds (Kluwer, 1989)

\bibitem{gunn78}	J.E. Gunn et al.,
Ap.J. 223 (1978) 1015

\bibitem{stecker78}	F.W. Stecker, Ap.J. 223(1978)1032

\bibitem{silk84}	J.Silk and M. Srednicki, Phys. Rev. Lett. 53 (1984) 624

\bibitem{bergstrom88}	L. Bergstr\"{o}m and H. Snellman, Phys. Rev. D37 (1988)
3737

\bibitem{rudaz89}	S. Rudaz, Phys. Rev. D39 (1989) 3549

\bibitem{bouquet89}	A. Bouquet, P. Salati and J. Silk, Phys. Rev. D40 (1989)
3168

\bibitem{weekes88}	T.C. Weekes, Phys. Rep. 160 (1988) 1

\bibitem{nilles84}	H.P. Nilles, Phys. Rep. 110 (1984) 1

\bibitem{nishimura80}	J.Nishimura et al., Ap.J. 238 (1980) 394

\bibitem{he91}	T.D.He and Q.Q. Bhu, Phys.Rev. D44 (1991) 2635.

\bibitem{mayer82}	H.A. Mayer-Hasselwander et al., Astr. and Astroph. 105 (1982)
164

\bibitem{bhat85}	Chaman L.Bhat et al., Nature 314 (1985) 511

\bibitem{bloemen89}	H.Bloemen, Ann.Rev.Astr.Astroph. 27 (1989)

\bibitem{bahcall80}	J. N. Bahcall and R.M. Soneira, Astrophys. J. Supp. 44
(1980) 73

\bibitem{caldwell81}	J.A.R. Caldwell and J. Ostriker, Ap. J. 251 (1981) 61

\bibitem{binney87}	J. Binney and S. Tremaine, ''Galactic Dynamics'' (Princeton
University Press, Princeton, 1987)

\bibitem{flores88}	R.A. Flores, Phys. Lett.B 215 (1988) 73

\bibitem{zeldovich80}	Ya.B. Zel'dovich et al, Sov. J. Nucl. Phys. 31 (1980) 664

\bibitem{clemens85}	D.P. Clemens, Ap. J. 295 (1985) 422

\bibitem{ipser87}	J.R. Ipser and P. Sikivie, Phys. Rev. D35 (1987) 3695

\bibitem{bengtsson87}	H.U. Bengtsson and T. Sjostrand, Computer Phys. Comm. 46
(1987) 43

\bibitem{kolb90}	E.W. Kolb and M.S. Turner, ''The early universe'',
(Addison-Wesley, New-York 1990)

\bibitem{rees86}	M.J. Rees, ''Baryonic dark matter'', 2nd ESO-CERN Symposium
(1986)

\bibitem{griest90}	K.Griest, M.Kamionkowski and M.S.Turner, Phys. Rev. D41
(1990) 3565

\bibitem{ellis84}	J.Ellis et al., Nucl. Phys. B238 (1984) 453

\bibitem{haber85}	H.E. Haber and G.L. Kane, Phys. Rep. 117 (1985) 75

\bibitem{griest88}	K. Griest, Phys. Rev. D38 (1988) 2357

\bibitem{olive89}	K.A. Olive and M. Srednicki, Phys. Lett. B230 (1989) 78

\bibitem{decamps90}	D. Decamps et al., ALEPH Collaboration, Phys. Lett. 244B
(1990) 541

\bibitem{alitti90}	J. Alitti et al., UA2 Collaboration, Phys. Lett. 235B (1990)
363

\bibitem{abe89}	F. Abe et al., CDF Collaboration, Phiys. Rev. Lett. 62 (1989)
1825ys. Rev. Lett. 62 (1989)
1825

\bibitem{barbieri88}	R. Barbieri and G.F. Giudice, Nucl. Phys. B306 (1988) 63
o: hep-ph@xxx.lanl.gov
Subject: put

\bibitem{giudice89}	G.F. Giudice and K. Griest, Phys.Rev. D40 (1989) 2549
\end{thebibliography}
\end{document}